\documentclass[aps,pra,twocolumn,showpacs,amsmath,amssymb, superscriptaddress]{revtex4-1}

\usepackage{graphicx}
\usepackage{dcolumn}
\usepackage{bm}
\usepackage{color}
\usepackage{comment}

\begin{document}

\preprint{APS/123-QED}

\title{Dispersion cancellation in high resolution two-photon interference 
\\ }

\author{Masayuki Okano}
\affiliation{%
Research Institute for Electronic Science, Hokkaido University,
Sapporo 001--0020, Japan
}%
\affiliation{%
The Institute of Scientific and Industrial Research, Osaka University,
Osaka 567--0047, Japan
}%

\author{Ryo Okamoto}
\affiliation{%
Research Institute for Electronic Science, Hokkaido University,
Sapporo 001--0020, Japan
}%
\affiliation{%
The Institute of Scientific and Industrial Research, Osaka University,
Osaka 567--0047, Japan
}%

\author{Akira Tanaka}
\affiliation{%
Research Institute for Electronic Science, Hokkaido University,
Sapporo 001--0020, Japan
}%
\affiliation{%
The Institute of Scientific and Industrial Research, Osaka University,
Osaka 567--0047, Japan
}%

\author{Shutaro Ishida}
\affiliation{%
Department of Electrical Engineering and Computer Science, 
Nagoya University, Nagoya 464--8603, Japan
}%

\author{Norihiko Nishizawa}
\affiliation{%
Department of Electrical Engineering and Computer Science, 
Nagoya University, Nagoya 464--8603, Japan
}%

\author{Shigeki Takeuchi}
\altaffiliation[Electronic address: ]{takeuchi@es.hokudai.ac.jp}
\affiliation{%
Research Institute for Electronic Science, Hokkaido University,
Sapporo 001--0020, Japan
}%
\affiliation{%
The Institute of Scientific and Industrial Research, Osaka University,
Osaka 567--0047, Japan
}%


\date{\today}

\begin{abstract}

The dispersion cancellation observed in Hong-Ou-Mandel (HOM) interference between frequency-entangled photon pairs has been the basis of quantum optical coherence tomography and quantum clock synchronization. Here we explore the effect of phase dispersion on ultranarrow HOM dips. We show that the higher-order dispersion, the line width of the pump laser, and the spectral shape of the parametric fluorescence have a strong effect on the dispersion cancellation in the high-resolution regime with several experimental verifications. Perfect dispersion cancellation with a linewidth of 3 $\mu$m is also demonstrated through 25 mm of water.

\end{abstract}

\maketitle


\section{Introduction}
 Two-photon interference (TPI) first demonstrated by Hong, Ou, and Mandel (HOM) \cite{Hong1987}, 
has become a universal concept in quantum optics. 
The dispersion cancellation observed in HOM interference 
between frequency-entangled photon pairs \cite{{Steinberg1992},{Steinberg1992-2}} 
is one of the most remarkable phenomena, 
and has been the basis of novel concepts: 
quantum optical coherence tomography (QOCT) \cite{{PhysRevA.65.053817},{PhysRevLett.91.083601}}
and quantum clock synchronization (QCS) \cite{PhysRevLett.87.117902}. 
Furthermore, the scientific interest in two-photon interference 
between frequency-entangled photon pairs is increasing rapidly, 
for example, in time-frequency entanglement measurement by weak measurements \cite{Hofmann2013} 
and multi-mode frequency entanglement \cite{Mikhailova2008}.

In these concepts and applications, it is critically important 
to realize high-resolution (``ultranarrow'' linewidth) HOM dips against phase dispersion, 
since the depth resolution of QOCT or the time-synchronization accuracy of QCS 
is determined by the linewidth of HOM dips. 
In more detail, the resolution of optical coherence tomography \cite{Huang1991}
using low-coherence interference (LCI) \cite{Max1999}
is highly limited by the group-velocity dispersion (GVD) \cite{{Drexler1999},{Fercher2001}}; 
a resolution of 20 $\mu$m is typical in ophthalmography 
without dispersion compensation \cite{Hitzenberger1999}. 
Since TPI can achieve better resolution than that of LCI, 
QOCT is expected to be an alternative to current OCT. 
However, in the past experimental test of dispersion cancellation, 
the linewidth of the HOM dip was limited to 19 $\mu$m \cite{PhysRevLett.91.083601}, 
which can still be well explained by the theory proposed in 1992 
\cite{{Steinberg1992},{Steinberg1992-2}}. 
Note that recently the HOM dip with a resolution of approximately 1 $\mu$m 
was demonstrated \cite{{Nasr2008},{Nasr2008-2}}; 
however, these experimental demonstrations 
were without any dispersive media in the optical paths.
 
The main motivation of this paper is 
to explore the dispersion cancellation in ``ultranarrow'' HOM dips.
It was conjectured that 
the nonzero pump-laser linewidth and the higher-order dispersion
may limit the dispersion cancellation in the high resolution TPI; 
however, these effects have not previously been studied well. 
Furthermore, in the former studies, 
there have been discrepancies in the resolution enhancement factor
of TPI to LCI. 
Some studies obtained a value of 
2 \cite{{PhysRevA.65.053817},{PhysRevLett.91.083601}}, 
while others obtained $\sqrt{2}$ \cite{Lavoie2009}. 

Here we discuss the generalized theory of 
TPI, which can explain all of these effects 
and go on to describe the rigorous testing of the theory, 
namely, the dispersion tolerance of high--resolution TPI. 
We show that the resolution enhancement factor 
depends on the spectral shape of the photons, 
changing from $\sqrt{2}$ for a Gaussian shaped spectrum 
to $2$ for a rectangular one. 
For the experimental testing of the theory, 
we constructed a hybrid LCI and TPI setup 
on which LCI and TPI experiments could be performed 
using the same parametric fluorescence. 
We also succeeded in observing the broadening of the TPI signal 
that occurs due to the nonzero pump-laser linewidth, 
which can be well explained by the proposed theory. 
From our results, 
it is apparent that we achieved ``perfect dispersion cancellation'' at high resolutions; 
the resolution of 3.0 $\mu$m was unchanged 
even when 25 mm of water, 
which corresponds to the diameter of the human eye, 
was placed in the optical path. 
In further testing we used 200 mm of water and observed a strong tolerance; 
the degradation of the resolution in TPI
was tiny (from 3.0 to 3.5 $\mu$m), 
while the resolution of LCI
degrades from 3.0 to 410 $\mu$m in theory.  
We believe our generalized theory and rigorous experimental proofs are beneficial 
not only to QOCT and QCS, but also to the basic understanding 
of multi-particle quantum interferences, 
which are the basis of quantum metrology and
quantum information science.

\section{A generalized theory of TPI}
Here we derive an equation to describe the HOM dips, 
taking into account the effects 
of the arbitrary spectral shape of the parametric fluorescence, 
linewidth of the pump laser, and higher-order dispersion of the media.

\begin{figure}[b]
\includegraphics[width=8.5cm]{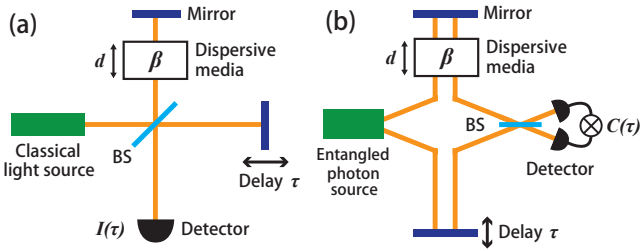}
\caption{\label{fig0} (color online).
Schematic of LCI (a) and TPI (b).
BS stands for beam splitter.
$\tau$ is the temporal delay and $\beta$ 
is the wavevector in a dispersive media of thickness $d$.}
\end{figure}

Before discussing TPI, we consider LCI.
Fig. \ref{fig0}(a) shows a scheme of the LCI.
The normalized interferogram $I(\tau)$ with a temporal delay $\tau$ between two arms 
is given by 
\begin{eqnarray}
I(\tau)=1+\textrm{Re} \{e^{-i\omega_0\tau}
\int d\Omega |f_{\textrm{o}}(\Omega)|^2 e^{-i\Omega \tau
+i2d\beta(\Omega)} \},
\label{OCTdis}
\end{eqnarray}
where $\Omega$ is the frequency deviation about the central frequency $\omega_0$,
$f_{\textrm{o}}$ is a normalized spectral probability amplitude \cite{Brezinski2006},
$d$ is the thickness of the media and 
$\beta(\Omega)=\beta^{(0)}+\beta^{(1)}\Omega+\frac{1}{2!}\beta^{(2)}\Omega^2+\cdots$ 
is the wavevector of the light in the dispersive media. 
$\beta^{(1)}$ is the inverse of the group velocity
and $\beta^{(2)}$ represents the GVD. 
Here we define the resolution $\Delta L$ by 
the full width at half maximum (FWHM) of the interferogram. 
Without any dispersive media, 
$\Delta L$ is the FWHM $W$ of the Fourier transform of the power spectrum 
$W(\textrm{FT}\{|f_{\textrm{o}}(\Omega)|^2\})$.
With a dispersive media, 
the degraded resolution $\Delta L^{\prime}$ due to the GVD is given by 
$\Delta L^{\prime}=
\Delta L (1+(2\sqrt{\ln 2}\times c \sqrt{\beta^{(2)}d}/\Delta L)^4)^{\frac{1}{2}}$
\cite{Brezinski2006}, where $\Delta L$ is the original resolution 
and $c$ is the speed of light in a vacuum.
Thus, $\Delta L^{\prime}$ has a minimum value of $\Delta L$ as seen from
$\Delta L_{th}\sim 2\sqrt{2\ln 2}\times c \sqrt{\beta^{(2)}d}$, 
which becomes zero when $\beta^{(2)}$ and $d$ are zero, 
as is the case when no dispersive media is present. 
The degradation becomes more significant
when $\Delta L$ is small (high resolution).
For example, 
the OCT resolution
is generally limited to approximately 20 $\mu$m due to the dispersion 
of the human eye ($\beta^{(2)}d\sim$720 fs$^2$) \cite{VanEngen1998}.

Next, we discuss the interferograms measured 
by a TPI setup (Fig. \ref{fig0}(b)). 
At first, the biphoton state produced by parametric down-conversion from a source 
can be described by 
\begin{eqnarray}
\lvert\Psi\rangle=\iint d\omega_{\textrm{s}} d\omega_{\textrm{i}}
f_{\textrm{q}}(\omega_{\textrm{s}}, \omega_{\textrm{i}})
\lvert\omega_{\textrm{s}}\rangle_{\textrm{s}}
\lvert\omega_{\textrm{i}}\rangle_{\textrm{i}},
\label{QOCTstatefull}
\end{eqnarray}
where 
$f_{\textrm{q}}(\omega_{\textrm{s}}, \omega_{\textrm{i}})$
is a normalized spectral probability amplitude of the biphoton and
$\lvert\omega_{\textrm{s(i)}}\rangle_{\textrm{s(i)}}$ 
represents the signal (idler) photon state with the frequency $\omega_{\textrm{s(i)}}$
\cite{Campos1990}.
The biphoton spectral amplitude $f_{\textrm{q}}$ can be given by 
\begin{eqnarray}
f_{\textrm{q}}(\omega_{\textrm{s}}, \omega_{\textrm{i}})
=\xi_{\textrm{p}}(\omega_{\textrm{s}},\omega_{\textrm{i}})
f_{\textrm{s}}(\omega_{\textrm{s}})
f_{\textrm{i}}(\omega_{\textrm{i}}),
\label{xi_q}
\end{eqnarray}
where $\xi_{\textrm{p}}(\omega_{\textrm{s}},\omega_{\textrm{i}})$
represents the phase-matching condition
and $f_{\textrm{s(i)}}(\omega_{\textrm{s(i)}})$ is 
a normalized spectral probability amplitude of the signal (idler) photons
\cite{Joobeur1994}.
The spectrum of each photon of the biphoton is determined
by a spatial mode where each photon resides.
Then the normalized coincidence count rate $C(\tau)$ of two output modes
with a temporal delay $\tau$ is given by 
\begin{eqnarray}
C(\tau)&=&1-\textrm{Re} \{\iint d\omega_{\textrm{s}} d\omega_{\textrm{i}}
f_{\textrm{q}}(\omega_{\textrm{s}}, \omega_{\textrm{i}})
f_{\textrm{q}}^*(\omega_{\textrm{i}}, \omega_{\textrm{s}}) \nonumber \\
&\times&
e^{i\phi(\omega_{\textrm{s}}, \omega_{\textrm{i}}, \tau)
+i\eta(\omega_{\textrm{s}}, \omega_{\textrm{i}})}\},
\label{QOCTeqfull}
\end{eqnarray}
where phase terms $\phi$ and $\eta$ account for 
the temporal delay $\tau$ and the dispersion $\beta$, respectively.
This is the generalized equation of a TPI interferogram, 
from which the dispersion tolerance at high resolutions can be deduced.
Based on this equation,
we discuss the resolution enhancement of TPI compared to LCI,
the degradation of the resolution due to the nonzero pump linewidth,
and the higher-order dispersion effect.

First we consider the resolution enhancement. 
For simplicity, we consider the case in which a pump laser is monochromatic. 
For a monochromatic pump with the frequency $\omega_{\textrm{p}}$, 
signal and idler photon frequencies are perfectly correlated 
due to the conservation of the energy 
$\omega_{\textrm{p}}=\omega_{\textrm{s}}+\omega_{\textrm{i}}$,
and the biphoton spectral amplitude is written as
\begin{eqnarray}
f_{\textrm{q}}(\omega_{\textrm{s}}, \omega_{\textrm{i}})=
\delta(\omega_{\textrm{p}}-\omega_{\textrm{s}}-\omega_{\textrm{i}})
f_{\textrm{s}}(\omega_{\textrm{s}})f_{\textrm{i}}(\omega_{\textrm{i}}),
\label{Delta}
\end{eqnarray}
where $\xi_{\textrm{p}}$ is a delta function.
Substituting Eq. (\ref{Delta}) into Eq. (\ref{QOCTstatefull}), 
the biphoton state is given by
\begin{eqnarray}
\lvert\Psi\rangle&=&
\int d\omega_{\textrm{s}}d\omega_{\textrm{i}}
\delta(\omega_p-\omega_s-\omega_i)
f_{\textrm{s}}(\omega_{\textrm{s}})f_{\textrm{i}}(\omega_{\textrm{i}})
\lvert\omega_{\textrm{s}}\rangle_{\textrm{s}}
\lvert\omega_{\textrm{i}}\rangle_{\textrm{i}} \nonumber \\
&=&
\int d\omega_{\textrm{s}}
f_{\textrm{s}}(\omega_{\textrm{s}})f_{\textrm{i}}(\omega_{\textrm{p}}-\omega_{\textrm{s}})
\lvert\omega_{\textrm{s}}\rangle_{\textrm{s}}
\lvert\omega_{\textrm{p}}-\omega_{\textrm{s}}\rangle_{\textrm{i}}\nonumber \\
&=&
\int d\Omega
f_{\textrm{s}}(\Omega)
f_{\textrm{i}}(-\Omega)
\lvert\frac{\omega_{\textrm{p}}}{2}+\Omega\rangle_{\textrm{s}}
\lvert\frac{\omega_{\textrm{p}}}{2}-\Omega\rangle_{\textrm{i}}\nonumber \\
&\equiv &
\int d\Omega
f_{\textrm{q}}(\Omega)
\lvert\frac{\omega_{\textrm{p}}}{2}+\Omega\rangle_{\textrm{s}}
\lvert\frac{\omega_{\textrm{p}}}{2}-\Omega\rangle_{\textrm{i}},
\label{Entangled}
\end{eqnarray}
where the signal (idler) photon frequency $\omega_{s(i)}$ is 
$\omega_p/2 \pm \Omega$ with the frequency deviation $\Omega$ and
$f_{\textrm{q}}(\Omega) \equiv f_{\textrm{s}}(\Omega)f_{\textrm{i}}(-\Omega)$.
Equation (\ref{Entangled}) is usually used as a description of 
the entangled biphoton state \cite{{Steinberg1992-2},{PhysRevLett.91.083601}}.
Then the TPI interferogram in Eq. (\ref{QOCTeqfull}) can be written by  
\begin{eqnarray}
C(\tau)
&=&1-\textrm{Re} \left\{\int d\Omega 
f_{\textrm{q}}(\Omega) f_{\textrm{q}}^*(-\Omega) e^{-i2\Omega \tau}\right\}\nonumber \\
&=&1-\textrm{Re} \left\{\int d\Omega 
|f_{\textrm{q}}(\Omega)|^2 e^{-i2\Omega \tau}\right\},
\label{QOCTeq}
\end{eqnarray}
where $\phi$ is $-2\Omega \tau$ and $\eta$ is zero.
The TPI resolution, which is the FWHM of the interferogram,
is half of the FWHM $W$ of the Fourier transform of the spectrum
$W(\textrm{FT}\{|f_{\textrm{q}}(\Omega)|^2\})/2$,
and the resolution enhancement factor $R_e$ is given by
\begin{eqnarray}
R_e=2\frac{W(\textrm{FT}\{|f_{\textrm{o}}(\Omega)|^2\})}
{W(\textrm{FT}\{|f_{\textrm{s}}(\Omega)|^2|f_{\textrm{i}}(\Omega)|^2\})}.
\label{R_q}
\end{eqnarray}
Now we discuss $R_e$ for the same bandwidth of 
LCI and TPI ($f_{\textrm{o}}$=$f_{\textrm{s,i}}$). 
Equation (\ref{R_q}) suggests that the resolution enhancement depends 
on the spectral shape of the photons.
$R_e$ is 2 for a rectangular shaped spectrum,
where $|f_{\textrm{s}}(\Omega)||f_{\textrm{i}}(\Omega)|$=$|f_{\textrm{s,i}}(\Omega)|$.
For a Gaussian shaped spectrum,
$R_e$ reduces to $\sqrt{2}$
because the biphoton spectral width is reduced by $\sqrt{2}$ 
compared to the single photon spectral width. 
These results suggest that the discrepancy observed in the previous reports 
\cite{{PhysRevA.65.053817},{PhysRevLett.91.083601},{Lavoie2009}}
is due to the difference in the spectral shape of the source used.

Next we consider the dispersion tolerance taking into account a nonzero pump linewidth.
When the pump laser has a nonzero linewidth, 
the biphoton spectrum is given by Eq. (\ref{xi_q}) and 
then the TPI interferogram in Eq. (\ref{QOCTeqfull}) 
with a dispersive media can be written as
\begin{eqnarray}
C(\tau)&=&1-\textrm{Re} \{\iint d\omega_{\textrm{s}} d\omega_{\textrm{i}}
|\xi_{\textrm{p}}(\omega_{\textrm{s}},\omega_{\textrm{i}})|^2
|f_{\textrm{s}}(\omega_{\textrm{s}})|^2
|f_{\textrm{i}}(\omega_{\textrm{i}})|^2 \nonumber \\
&\times&
e^{-i(\omega_{\textrm{s}}-\omega_{\textrm{i}})\tau
+i2d(\beta(\omega_{\textrm{s}})-\beta(\omega_{\textrm{i}}))}\},
\end{eqnarray}
where each photon spectral amplitude is assumed to be the same 
($f_{\textrm{s}}=f_{\textrm{i}}$).
For a monochromatic pump, 
$\eta$ is $4d\beta^{(1)}\Omega$ and the GVD $\beta^{(2)}$ is completely canceled
within the second-order dispersion approximation, indicating
a perfect dispersion tolerance.
However, the dispersion tolerance becomes imperfect for a pump with a certain nonzero linewidth.
When $\xi_{\textrm{p}}$ is the Gaussian of $\omega_{\textrm{p}}=\omega_{\textrm{s}}+\omega_{\textrm{i}}$
with the bandwidth $\Delta\omega_p$
and $f_{\textrm{s(i)}}$ is also Gaussian
with the bandwidth $\Delta \omega_{s(i)}$,
the original resolution $\Delta L$ degrades to
$\Delta L^{\prime}=(1+\left(\Delta \omega_{\textrm{s,i}} \Delta \omega_{\textrm{p}} 
\beta^{(2)}d / 8\sqrt{2}\ln 2 \right)^2)^\frac{1}{2}\Delta L$ 
\cite{Resch2009},
where $\beta^{(2)}$ cannot be canceled.
The degradation becomes more significant
when the pump linewidth is larger. 

Finally, we consider the higher-order dispersion effect.
For higher-order approximation of dispersion in the media,
the interferogram with a monochromatic pump can be written from Eq. (\ref{QOCTeqfull}) as 
\begin{eqnarray}
C(\tau)&=&1-\textrm{Re} \{\int d\Omega 
|f_{\textrm{q}}(\Omega)|^2 \nonumber \\
&\times& e^{-i2\Omega \tau+i4d(\frac{\beta^{(3)}}{3!}\Omega^3+\cdots)}\},
\label{QOCThigh}
\end{eqnarray}
where the low-order dispersion $\beta^{(1)}$ is omitted. 
The third-order dispersion $\beta^{(3)}$ is not canceled
as all odd-order dispersion. 
The asymmetric phase dispersion with the cubic dependence $\beta^{(3)}\Omega^3$
makes interferograms asymmetric \cite{Okamoto2006}.
Though the higher-order dispersion is small in general cases,
the contribution is not negligible
at high resolutions due to the large frequency component $\Omega$. 

\begin{figure}[b]
\includegraphics[width=8.5cm]{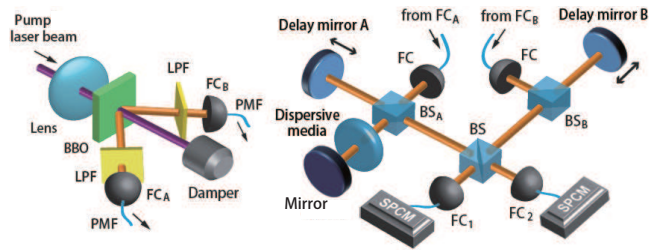}
\caption{\label{fig1} (color online).
Schematic of the experimental setup,
showing the light source (left) 
and the hybrid LCI/TPI interferometer (right).}
\end{figure}

\section{The hybrid LCI/TPI experimental setup}
To experimentally test the generalized theory for high resolution TPI,
we constructed a hybrid experimental setup 
for both LCI and TPI as illustrated in Fig. \ref{fig1}.
The cw pump-laser beam 
(wavelength, 404 nm; linewidth, $\sim$100 kHz; power, 100 mW)
is focused at a type-I phase-matched $\beta$-barium-borate (BBO) crystal, 
cut by a long-pass filter and stopped by a damper.
Parametric fluorescence occurs as frequency-entangled biphotons are generated
with a center wavelength of 808 nm and a bandwidth of 75 nm.
For LCI experiments,
photons transferred through a polarization-maintaining fiber 
from FC$_{\textrm{A}}$ serve as classical lights.
The optical delay $c\tau$ is determined by a delay mirror \textit{A}
and a dispersive media can be set in front of the other mirror.
Interfered light is detected 
by a single-photon-counting module (SPCM, Perkin Elmer, SPCM-AQRH-14).
For the TPI experiments,
signal photons from the FC$_{\textrm{A}}$ travel 
through the dispersive media,
and idler photons from the FC$_{\textrm{B}}$ are reflected at the delay mirror \textit{B} with an optical delay $c\tau$.
The coincidence count rate at the FC$_{\textrm{1,2}}$
is measured by two SPCMs.

\section{Results and discussion}
\subsection{Spectral-shape dependence of the resolution enhancement}
\begin{figure}[b]
\includegraphics[width=8.5cm]{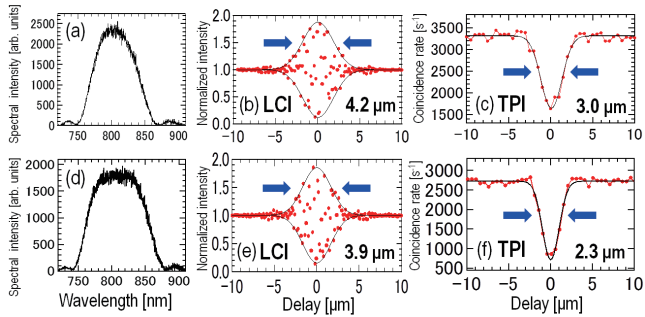}
\caption{\label{fig2} (color online).
Experimental LCI and TPI interferograms 
for two spectral shapes:
For the Gaussian shaped spectrum (a),
the 4.2-$\mu$m LCI (b) 
and the 3.0-$\mu$m TPI (c) widths
show a resolution enhancement factor $R_e$ of 1.4;
for the trapezoidal shaped spectrum (d),
the 3.9-$\mu$m LCI (e) 
and the 2.3-$\mu$m TPI (f) widths
show a value of 1.7 for $R_e$.
The experimental data (red dots) and
Gaussian fit curves (black solid line) are plotted.
The integration time is 1 s per point in Figs. (b), (c), (e) and (f).
}
\end{figure}

First, we verified the theory for the resolution enhancement factor $R_e$.
For a Gaussian shaped spectrum (Fig. \ref{fig2}(a)),
experimental LCI and TPI interferograms show FWHMs of 
4.2$\pm$0.1 (Fig. \ref{fig2}(b)) 
and 3.0$\pm$0.1 $\mu$m (Fig. \ref{fig2}(c)), respectively. 
Interferograms are expressed in units of the optical delay $c\tau/2$ 
to represent the physical displacement of the delay mirrors.
We used a Gaussian fit to measure the FWHMs of the interferograms.
The obtained $R_e$ was 1.4$\pm$0.1, 
which is consistent with the theoretical factor of $\sqrt{2}$. 
We then used a trapezoidal-shaped spectrum (Fig. \ref{fig2}(d))
produced by controlling the tilting angle of the BBO crystal.
The $R_e$ factor of 1.7$\pm$0.1 
obtained from the 3.9$\pm$0.1-$\mu$m LCI (Fig. \ref{fig2}(e))
and the 2.3$\pm$0.1-$\mu$m TPI (Fig. \ref{fig2}(f)) FWHMs
is also fairly consistent with the theoretical value of 1.69.
Thus, the experimental results supported our proposed theory.

\subsection{Perfect dispersion cancellation}
\begin{figure}[tb]
\includegraphics[width=8.0cm]{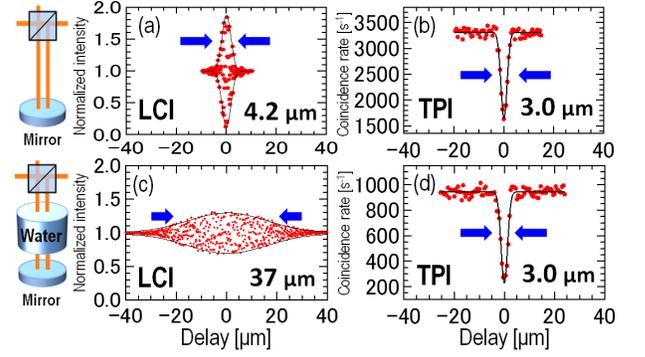}
\caption{\label{fig3}(color online).
Experimental LCI and TPI interferograms 
for the Gaussian shaped spectrum:
The 4.2-$\mu$m LCI (a) and 
3.0-$\mu$m TPI (b) widths
without any media and
the broadened 37-$\mu$m LCI (c) 
and unchanged 3.0-$\mu$m TPI (d) widths
using 25 mm of water as a dispersive media.
The integration time is 1 s per point in Figs. (a)--(d).}
\end{figure}

Next we performed perfect dispersion tolerance
with a narrowband pump laser.
A 25-mm-thick volume of water, which is about the diameter of the human eye, 
was enclosed by two 1-mm-thick BK7 glass plates 
and placed as the dispersive media. 
Dispersion tolerance for a water is important
in particular for the QOCT application.
LCI resolutions of 3.0 and 4.2 $\mu$m over the threshold
($\Delta L_{th}\sim$ 20 $\mu$m)
are expected to significantly degrade to 55 and 37 $\mu$m, respectively.
Actually, the LCI interferogram 
broadened to 36.8$\pm$0.4 $\mu$m (Fig. \ref{fig3}(c)) 
from the original resolution of 4.2 $\mu$m
(Fig. \ref{fig2}(b) and Fig. \ref{fig3}(a)).
In contrast, the TPI interferogram was unchanged 
with a width of 3.0$\pm$0.1 $\mu$m (Fig. \ref{fig3}(d))
from the original resolution of 3.0 $\mu$m
(Fig. \ref{fig2}(c) and Fig. \ref{fig3}(b)).
Thus, perfect dispersion tolerance has been successfully demonstrated.

\subsection{Degradation due to the nonzero  pump linewidth}
\begin{figure}[b]
\includegraphics[width=7.0cm]{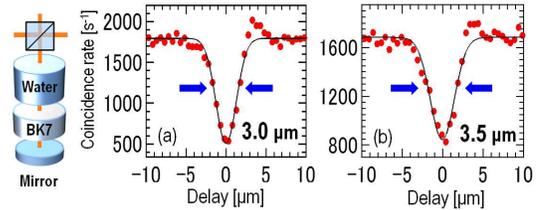}
\caption{\label{fig3_2} (color online).
TPI interferograms with dispersive media
(25 mm of water and a 20-mm-thick BK7 plate) 
using narrowband (100 kHz) (a) and
broadband (1 THz) (b) pumping--laser linewidths.
The nonzero pump linewidth degrades 
the resolution from 3.0 to 3.5 $\mu$m.
The integration time is 1 s per point in Figs. (a) and (b).
}
\end{figure}

To test the effect of the nonzero  pump linewidth, 
the 25 mm of water and a 20-mm-thick BK7 plate were used,
which gave a total GVD delay $\beta^{(2)}d$ of 1700 fs$^2$.
For the narrowband pump laser (linewidth$\sim$100 kHz),
complete dispersion cancellation was achieved
with an unchanged FWHM of 3.0$\pm$0.1 $\mu$m (Fig. \ref{fig3_2}(a)).
In contrast, for a broadband pump laser (linewidth$\sim$1 THz, a laser diode),
the FWHM broadened to 3.5$\pm$0.2 $\mu$m (Fig. \ref{fig3_2}(b))
with a degradation factor of 1.2$\pm$0.1.
This broadening is close to the theoretical value of 1.17,
meaning our proposed theory has been verified.

\subsection{Higher-order dispersion effect}
\begin{figure}[tb]
\includegraphics[width=7.0cm]{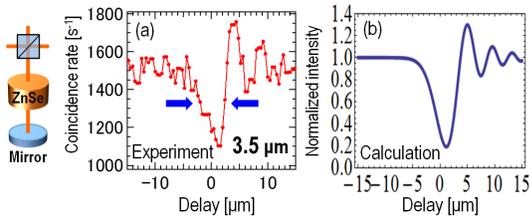}
\caption{\label{fig4} (color online).
The experimental TPI interferogram 
with a 5-mm-thick ZnSe plate as a highly dispersive media (a)
and a theoretical plot considering the third-order dispersion (b).
The integration time is 1 s per point in Fig. (a).}
\end{figure}

Finally, we tested the higher-order dispersion effect. 
A 5-mm-thick zinc selenide (ZnSe) plate 
($\beta^{(2)}\sim$1,000 fs$^2$/mm, $\beta^{(2)}d\sim$5,000 fs$^2$ \cite{bass2009handbook})
was used as a highly dispersive media.
$\beta^{(2)}d$ corresponds to the dispersion caused by 200 mm of water
which is expected to result in the LCI resolution degrading 
from 3.0 to 410 $\mu$m.
Even with this large GVD,
the TPI interferogram maintains a resolution of 3.5-$\mu$m FWHM
by an improvement of over 120 times (Fig. \ref{fig4}(a)).
The tiny degradation can be explained by the third-order dispersion effect.
A theoretically simulated TPI interferogram 
considering the third-order dispersion
($\beta^{(3)}\sim$870 fs$^3$/mm, $\beta^{(3)}d\sim$4,400 fs$^3$)
successfully reproduced the experimental result (Fig. \ref{fig4}(b)).
The sign of the asymmetry is determined by the sign of $\beta^{(3)}$,
and the oscillation comes from the high-frequency component $|\Omega|$.
A strong dispersion tolerance with a highly dispersive media has been observed, 
and the higher-order dispersion effect has been demonstrated.  

The resolution degradation factors due to the nonzero pump linewidth 
and the higher-order dispersion 
are both theoretically estimated to be 1.0 and nonobservable 
when the the resolution of 19 $\mu$m \cite{PhysRevLett.91.083601} is assumed. 
This fact suggests that 
these effects have become able to be observed 
due to our high-resolution (3 $\mu$m) two-photon interference experiment 
(the factors are 1.2 and 1.3, respectively, in our experiment.)

It should be also mentioned that 
the chirped-pulse interferometry \cite{{Lavoie2009},{Kaltenbaek2008}} 
and the phase-conjugate OCT \cite{{Erkmen2006},{LeGouet2010}} 
are also promising alternatives to QOCT. 
In these schemes, 
the photon flux of the light source can be much higher compared to QOCT, 
but the realized bandwidth of the correlated light source 
seems much broader in QOCT \cite{Tanaka2012}. 
Research using these alternative approaches 
will have beneficial positive mutual feedback for further investigations.

\section{Conclusion}
In conclusion, 
we investigated the dispersion tolerance of high-resolution HOM interference
and clarified all effects of the spectral shape of the source, 
the linewidth of the pump laser, 
and the higher-order dispersion 
using a generalized equation
in both theoretical and experimental sides at high resolutions.
We have found that the discrepancy in the resolution enhancement factor 
is due to the difference in the spectral shapes of sources
and experimentally verified the spectral dependence. 
We have also demonstrated perfect dispersion tolerance 
at 3-$\mu$m resolution by using a narrowband pump laser 
and observed a nonperfect dispersion tolerance 
for a nonzero pump-laser linewidth. 
We further demonstrated a strong dispersion tolerance 
in highly dispersive media and observed the higher-order dispersion effect.
Our results are important for QOCT and QCS 
and also offer important insights into
quantum information science and technology 
based on multi-particle quantum interferences.

\section*{Acknowledgments}
This work was supported in part by 
the Japan Science and Technology Agency CREST project;
the Quantum Cybernetics project of the Japan Society for the Promotion of Science (JSPS); 
the Funding Program for World-Leading Innovative R \verb|&| D on Science and Technology Program of JSPS;
Grant-in-Aids from JSPS;
the Project for Developing Innovation Systems of the Ministry of Education, Culture,
Sports, Science, and Technology;
the Network Joint Research Center for Advanced Materials and Devices; 
the Research Foundation for Opto-Science and Technology; and 
the Global COE program.
A.T. acknowledges the JSPS Research Fellowship program.

\bibliography{QOCTbib}

\end{document}